\documentclass[twocolumn,showpacs,amsmath,amssymb]{revtex4}
\usepackage{graphicx}
\usepackage{dcolumn}
\usepackage{bm}
\preprint{APS/123-QED}
\begin{document}

\title{Ultimate field-free molecular alignment by
combined adiabatic-impulsive field design}
\author{S.~Gu\'erin}
\email{sguerin@u-bourgogne.fr}
\author{A. Rouz\'ee}
\author{E. Hertz} \affiliation{Institut Carnot de Bourgogne, UMR
5209 CNRS, Universit\'e de Bourgogne, BP 47870, 21078 Dijon,
France}

\begin{abstract}
We show that a laser pulse designed as an adiabatic ramp followed
by a kick allows one to reach a perfect postpulse molecular
alignment, free of saturation. The mechanism is based on an
optimized distribution of the energy between a weakly efficient
but non saturating adiabatic ramp and an efficient but saturating
impulsive field.
 Unprecedent degrees of alignment are predicted using
state-of-the-art pulse shaping techniques and non-destructive
field intensities. The scheme can be extended to reach high
degrees of orientation of polar molecules using designed
half-cycle pulses.

\end{abstract}

 \maketitle

Control of the angular distribution of molecules in space features
major applications in physics and chemistry (see
\cite{Seideman_review} for a review). In this quest, an important
issue concerns the production of a high degree of field-free
alignment induced by non-destructive fields, especially for
non-cold molecules \cite{Stap,Vrakking,DeNalda,HRGLF}. It is
well-known that an intense non-resonant impulsive laser field
(i.e. of duration much shorter than the classical rotational
motion of the molecule) leads to field-free alignment, appearing
as periodic revivals. While the degree of alignment can be
improved with an increasing field intensity, it is however limited
by (i) destructive competing processes such as ionization and (ii)
an intrinsic saturation as obtained in the model of the rigid
rotor driven by a laser kick. Trains of kicks have been proposed
to shift to higher values this intrinsic saturation of alignment
\cite{Rabitz} and also in the context of molecular orientation
where the laser fields are replaced by Half-cycle pulses (HCP)
\cite{reaching,Dion}. However, such trains of pulses also show
plateaus of saturation. In practice, intense and generally
ionizing kicks will be required to reach a high degree of
alignment, in particular for non-cold molecules. An alternative
solution \cite{Seideman,Stolow2003} consists in truncating an
adiabatic pulse to produce a so-called switched wavepacket
exhibiting revivals of alignment. These revivals are free of
saturation as can be intuitively understood from the structure of
the field-induced double well potential that can be made as deep
as desired. However, besides the complexity to produce such pulse
shapes with sufficiently high intensities, this truncated
adiabatic method suffers from its energetic inefficiency to reach
a high degree of alignment.

In this paper, we show that designing a laser pulse combining both
ingredients, i.e. made of a combination of an adiabatic ramp
followed by a kick, allows one to overcome the intrinsic
saturation in a robust way, and thus to reach unprecedent degrees
of alignment for cold and moderately hot molecules, using
non-destructive fields. This strategy generalizes an optimization
procedure reported in Ref. \cite{HRGLF} to maximize the alignment
of the O$_2$ molecule at a given pulse energy using pulse-shaping
techniques.

The problem of optimization is considered as follows. For a given
fluence of the kick (corresponding to the laser energy per surface
unit), whose peak intensity and duration are limited respectively
by the ionization of the molecule and by the conditions of the
impulsive regime, we determine the minimum intensity and duration
of the adiabatic ramp, maximizing the subsequent field-free
alignment.

When a linear molecule is subjected to a non-resonant linearly
polarized laser pulse, the effective dressed Hamiltonian reads
\cite{Fried95}
\begin{equation}
\label{Ham} H=BJ^{2}-\frac{1}{4}\mathcal{E}^{2}(t)
\Delta\alpha\cos^{2}\theta
\end{equation}
with $B$ the molecular rotational constant, $\Delta\alpha>0$ the
anisotropic polarizability, $\mathcal{E}(t)$ the laser field
envelope, and $\theta$ the angle between the laser polarization
axis and the molecular axis. Coupling terms independent of
$\theta$ have been omitted since they do not affect the alignment,
which is quantified through the observable
$\langle\cos^2\theta\rangle$. The molecule is effectively subject
to a double-well potential of the form $-\cos^{2}\theta$, with
minima at $\theta=0$ and $\theta=\pi$ corresponding to the
magnetic quantum number $M=0$.
The condition for adiabatic transport that allows one to connect
the initial state, say the lowest one $|J=0\rangle$ for $T=0$, to
the lowest pendular state $|\widetilde J=0\rangle$, eigenvector of
the Hamiltonian (\ref{Ham}), is estimated with respect to the
smallest detuning $6B$ of the two-photon process between the
rotational states:
$\tau_a\gg\hbar/6B,$
$\tau_a$ characterizing the pulse duration. The adiabatic maximum
alignment is given by \cite{Fried95}
\begin{equation}
\label{gamma} \max_t\langle\cos^{2}\theta\rangle_a\sim
1-1/\sqrt{\gamma},\quad \gamma=\mathcal{E}_a^{2}\Delta\alpha/4B
\end{equation}
in the limits of low temperature $T$, i.e. $kT/B\ll1$, and of high
field regime $\gamma\gg1$.  If we consider smooth pulses, such as
Gaussian pulses of shape
$\mathcal{E}_a\sqrt{\Lambda(t)}=\mathcal{E}_a
e^{-2\log{2}(t/\tau_a)^{2}}$ (with $\mathcal{E}_a$ the peak
amplitude of the field and $\tau_a$ the full width at half maximum
of the intensity), the adiabatic regime is already quite well
achieved for $\tau_a \sim T_\text{rot}\equiv\pi\hbar/B$. As a
function of the fluence of a Gaussian ramp at its peak value
$F_a=\mathcal{E}_a^{2}\int_{-\infty}^0 dt
\Lambda(t)\sim\mathcal{E}_a^{2}\tau_a/2\sim
3\mathcal{E}_a^{2}\hbar/2B$, we obtain
$\langle\cos^{2}\theta\rangle_a\sim
1-\sqrt{6\hbar/F_a\Delta\alpha}$.
 This shows that the adiabatic alignment
does not saturate but grows very slowly toward 1 as a function of
the pulse fluence. For instance, we would obtain typically for the
CO$_2$ molecule $\langle\cos^{2}\theta\rangle_a\approx 0.987$ and
$\langle\cos^{2}\theta\rangle_a\approx 0.913$, at temperature
respectively $T=0$ and $T=30$ K, by a truncated adiabatic pulse of
peak intensity $I=100$ TW/cm$^2$ and of width $\tau_a\approx 50$
ps. That corresponds to a pulse fluence above the saturation of
the ionization.

On the other hand, the alignment in the impulsive regime, i.e.
associated to the Hamiltonian in the impulsive approximation
$H_{\text{imp.}}=BJ^{2}-2\hbar\zeta\delta(t)\cos^{2}\theta$,
with $\delta(t)$ the Dirac distribution, is characterized by the
dimensionless quantity
\begin{equation}
\zeta=F_k\Delta\alpha/8\hbar,\quad F_k=\mathcal{E}_k^{2}\int
dt\Lambda(t)
\label{zeta}%
\end{equation}
with $F_k$ the fluence of the laser kick. For $T=0$, the maximum
alignment grows as a function of $\zeta$ as
\begin{equation}
\max_t\langle\cos^{2}\theta\rangle_k\sim
c_s-(c_s-1/3)e^{-\zeta^{3/2}},
\end{equation}
until reaching the saturation at
$c_s\equiv\langle\cos^{2}\theta\rangle_k^{(\text{sat.})}\sim
0.92.$
The efficiency of the kick is shown through the exponential growth
of the alignment.

The mechanism we propose consists in combining a saturating but
efficient laser kick with a non-saturating adiabatic ramp. The
adiabatic ramp prepares a well-aligned state, which can be
efficiently enhanced by the subsequent kick. The optimum
compromise between the two fluences is more precisely established
below. We first show the efficiency of alignment for the ideal
conditions independent of the particular molecule (cold molecule,
rigid rotor model, adiabatic regime for the ramp, and kick for the
impulsive part of the pulse) and show how this efficiency survives
when deviations from these conditions are considered.

\begin{center}
\begin{figure}[t]
\includegraphics[scale=0.7]{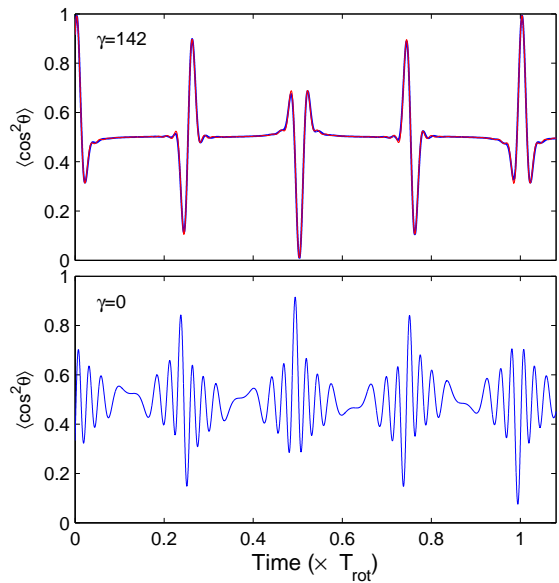}
\caption{(Color online) Upper frame: Alignment determined through
the observable $\langle\cos^2\theta\rangle(t)$ (thick line) and
$\langle\psi_{13}|\cos^2\theta|\psi_{13}\rangle(t)$ with the
relevant target state $|\psi_{13}\rangle$ (thin line), calculated
for $\zeta=11$ (in the impulsive approximation), $\gamma=142$ and
temperature $T=0$ as a function of the normalized time
$t/T_\text{\scriptsize rot}$. This corresponds to non-ionizing
intensities $I_k=50$ TW/cm$^2$ (with $\tau_k=100$ fs) and
$I_a=2.5$ TW/cm$^2$ for the CO$_2$ molecule
($T_\text{rot}\approx42.8$ ps). The maximum alignment is
$\max_t\langle\cos^2\theta\rangle\approx0.993$. Lower frame:
$\langle\cos^2\theta\rangle(t)$ in the same condition as above but
with $I_a=0$.} \label{c2tt}
\end{figure}
\end{center}

The upper frame of Fig. 1 shows a typical signal
$\langle\cos^2\theta\rangle(t)$ (of period $T_\text{rot}$),
calculated at $T=0$, when the respective adiabatic and impulsive
regimes are met (for an ideal model of a rigid rotor and with the
impulsive approximation for the kick). The kick defines the origin
of time $t=0$. The predicted maximum degree of alignment
$\max_t\langle\cos^2\theta\rangle\approx0.993$ is of unprecedent
efficiency under non-destructive field intensities for most of the
molecules. It is obtained right after the kick (or equivalently
after one full period $T_\text{\scriptsize rot}$), precisely at a
time $t_{\max}$ which gets closer to 0 for higher intensities. In
the conditions of the upper frame of Fig. 1, the maximum is
located at $t_{\max}\approx0.0037T_\text{rot}$. The signal
exhibits a remarkably simple form of four revivals in a period
with no noticeable oscillations in-between.
This indicates a perfect rephasing of the wavepacket at the
occurrence of the revivals. This strongly contrasts when kicks are
applied without adiabatic ramp, as shown in the lower frame of
Fig. 1.

We have compared the alignment with the one obtained with an ideal
target state. The target state is defined as the state
$|\psi_N\rangle$ that leads to the maximum alignment in the
truncated Hilbert subspace spanned by the first $N$ rotational
states \cite{reaching}. It corresponds to the eigenstate of
$\cos^2\theta_N$ in this subspace associated to the maximum
eigenvalue. We obtain, in the conditions of the upper frame of
Fig.1, the projection on the target state
$|\langle\psi_{N=13}|\phi(t_{\max})\rangle|^2\approx 0.996$ at the
time corresponding to the maximum of alignment, with $\phi(t)$ the
state at time $t$. We determine the dimension $N$ of the target
state taking the value that leads to the maximum projection
$|\langle\psi_{N}|\phi(t_{\max})\rangle|$. The alignment that
would be obtained ideally with this target state
$\langle\psi_{13}|\cos^2\theta|\psi_{13}\rangle(t)$ is almost
undistinguishable from the one obtained with
$\langle\cos^2\theta\rangle(t)\equiv\langle\phi(t)|\cos^2\theta|\phi(t)\rangle$.

\begin{center}
\begin{figure}[b]
\includegraphics[scale=0.7]{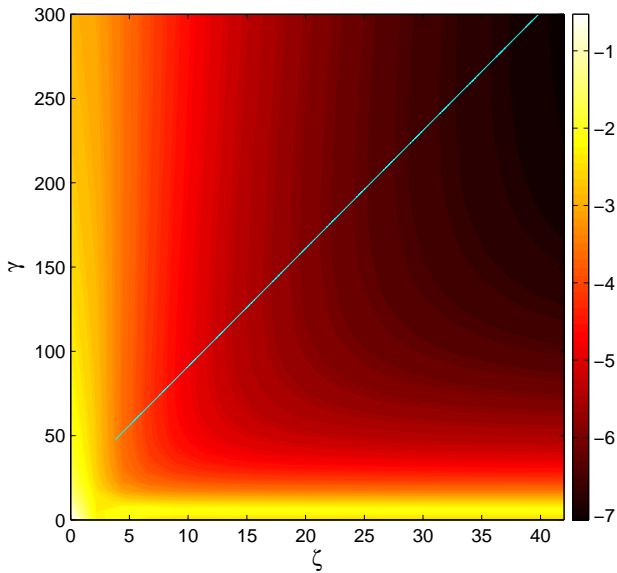}
\caption{(Color online) Contour plot of
$\log(1-\max_t\langle\cos^2\theta\rangle)$ as a function of
$\zeta$ (in the impulsive approximation) and $\gamma$ for
temperature $T=0$. The straight line corresponds to the value of
$\gamma$ that gives the maximum alignment for a given $\zeta$
(which exists only for $\zeta\gtrsim3.8$).} \label{Cont_T0}
\end{figure}
\end{center}

Figure 2 displays a contour plot of the maximum postpulse
alignment (in a logarithmic scale) when the short and the
adiabatic pulses are combined, as a function of the parameters
$\zeta$ and $\gamma$ (and thus independently of the chosen
molecule). The adiabatic pulse dramatically enhances the alignment
at a given $\zeta$ and vice versa. For a given $\zeta\gtrsim3.8$
(which corresponds to the saturation value for $I_a=0$), there is
a value for $\gamma$ giving the best alignment. This optimum is
shown by the straight line of equation
$\gamma_{\text{opt}}\approx7\zeta+21$. On this optimum line,
$\max_t\langle\cos^2\theta\rangle$ approaches exponentially 1 as
$1-ae^{-b\sqrt{\zeta}+c\zeta}$, with $a\approx0.20$,
$b\approx1.2$, and $c\approx0.053$. Larger values of $\gamma$ are
shown to give a slightly smaller degree of alignment, that can be
considered as practically constant (for not too large values of
$\gamma$). One can improve the degree of alignment as much as
desired by increasing $\zeta$ and $\gamma$ in the direction of the
straight line. For a chosen $\zeta$, mainly responsible of the
ionization, one has to take the intensity as low as possible by
choosing the largest duration satisfying the impulsive regime.

Instead of a single kick, one can use a train of kicks centered at
the rotational periods. The adiabatic ramp intensity has to be
adapted in this case to $\gamma_{\text{opt}}=7\sum\zeta+21$, i.e.
to the sum of the $\zeta$'s of the kicks.

We have checked that this strategy can be extended for non-cold
molecules. In this case, the contour plot of Fig. 2 has the same
qualitative features. As expected, a similar degree of alignment
requires higher $\zeta$ and $\gamma$ for higher temperatures. The
line $\gamma_{\text{opt}}(\zeta)$ from where on the efficiency is
optimal is of larger slope and larger ordinate at origin for
larger temperatures. Even when the optimum cannot be reached in
practice, we already obtain high degrees of alignment. For
instance, for the normalized temperature $kT/B=50$, with even $J$
considered in the thermal distribution (corresponding to
$T\approx30$ K for the CO$_2$ molecule), we obtain the unprecedent
degree of alignment $\max_t\langle\cos^2\theta\rangle\approx
0.945$ for $\zeta=22$ and $\gamma=568$ in the impulsive regime
(corresponding to non-destructive intensities $I_a\approx10$
TW/cm$^2$ and $I_k\approx50$ TW/cm$^2$ with $\tau_k=200$ fs for
the CO$_2$ molecule, see below the small correction to the degree
of alignment when the impulsive approximation is not considered).
Other initial thermal distributions lead to similar degrees of
alignment, which implies that our strategy is applicable to a wide
range of molecules and temperatures.



In practice we have to consider the short pulse with a finite
duration and thus to quantify its effect on the degree of
alignment with respect to the ideal impulsive approximation. In
general, deviations from the impulsive regime at a given fluence
reduce the degree of alignment. In the conditions of the upper
frame of Fig. 1 (but not in the impulsive approximation), we have
calculated the maximum degrees of alignment
$\max_t\langle\cos^2\theta\rangle\approx0.988$ and
$\max_t\langle\cos^2\theta\rangle\approx0.991$ for respectively
$\tau_\text{FWHM}=0.005T_\text{rot}$ and
$\tau_\text{FWHM}=0.0025T_\text{rot}$  (corresponding respectively
to $\tau_\text{FWHM}=200$ fs and $\tau_\text{FWHM}=100$ fs for the
CO$_2$ molecule). In these cases, the maximum degree of alignment
is very close to $\max_t\langle\cos^2\theta\rangle\approx0.993$
obtained in the impulsive regime. The degree of alignment will be
more reduced with respect to the one obtained in the impulsive
approximation for higher temperature, since the impulsive regime
will be more difficult to be satisfied for initial conditions with
higher $J$.

\begin{center}
\begin{figure}[b]
\includegraphics[scale=0.7]{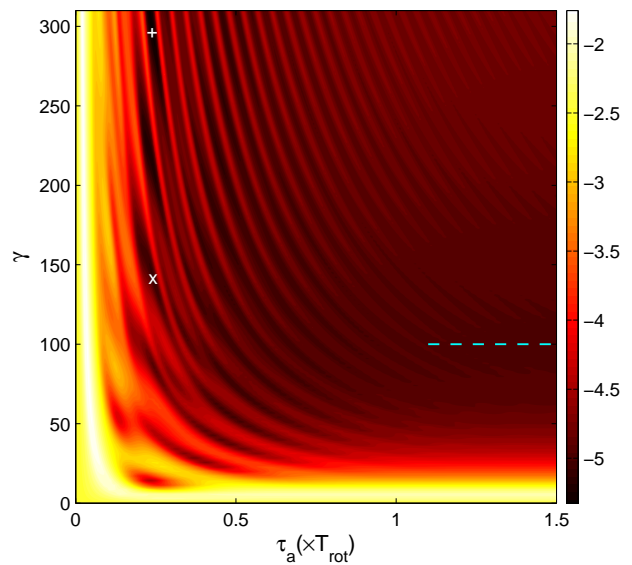}
\caption{(Color online) Contour plot of
$\log(1-\max_t\langle\cos^2\theta\rangle)$ as a function of the
normalized duration of the ramp $\tau_a/T_\text{rot}$ and $\gamma$
for temperature $T=0$ and $\zeta=11$ (in the impulsive
approximation). The straight dashed line corresponds to the
adiabatic regime in the conditions of the optimum straight line of
Fig. 2. The ``x'' and the ``+'', located at $\tau_a\approx
T_\text{rot}/4$ (for both) and respectively $\gamma=142$,
$\gamma=297$, indicate examples of very high degrees of alignment
(respectively $\max_t\langle\cos^2\theta\rangle\approx0.994$ and
$\max_t\langle\cos^2\theta\rangle\approx0.995$) for a ramp of
reduced duration with respect to an adiabatic ramp.}
\label{Cont_T0_Ia_Tr}
\end{figure}
\end{center}

In a next step, we study the deviations from the strict adiabatic
limit for the ramp. Figure 3 shows that they lead to a an
oscillating maximum of alignment from $\tau_a\sim T_\text{rot}/4$
up to the adiabatic regime. Specific ramp of intensities $\gamma$
and durations $\tau_a$ (as the ones indicated in Fig. 3) can thus
be found to induce the same degree of alignment as in the
completely adiabatic case (or even very slightly improving it). As
will be shown, this will benefit its practical implementation. We
have found that the duration of the ramp can be shortened,
allowing the dynamics to be non-adiabatic with respect to the
transition $J=0/J=2$, while approximately keeping adiabaticity
with respect to the upper transitions.
This strategy, that we will refer to as a combined
partially-adiabatic impulsive strategy, allows one to reach a
target state of higher dimension (thus of shorter duration) with
respect to the strategy using a strict adiabatic ramp. We have
indeed obtained
$|\langle\psi_{N=16}|\phi(t_{\max})\rangle|^2\approx 0.98$ and
$|\langle\psi_{N=17}|\phi(t_{\max})\rangle|^2\approx 0.99$ in the
conditions marked in Fig. 3 respectively by ``x'' and ``+''. This
combined partially-adiabatic impulsive strategy is less robust
with respect to the ramp shape, amplitude and duration as shown in
Fig. 3. But, we have checked that it is quite robust with respect
to the temperature, since it involves thermal initial conditions
of preserved adiabaticity with $J>2$. Figure 3 keeps indeed the
same features when a moderate temperature is considered.

No better alignment has been found by varying other parameters
(except of course the fluence of the short pulse, which is kept
fixed).

\begin{center}
\begin{figure}[t]
\includegraphics[scale=0.85]{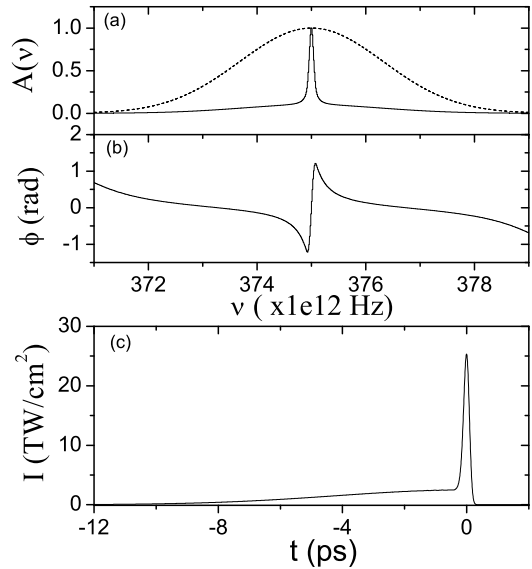}
\caption{Tailored spectral amplitude $A$ (upper frame, full line)
and phase $\phi$ (middle frame) from an inital 200 fs Fourier
transform limited pulse (upper frame, dashed line) to generate the
required field amplitude (lower frame)
with $I_k=25$ TW/cm$^2$, $\tau_k=200$ fs ($\zeta\approx 11$),
$I_a=2.5$ TW/cm$^2$ ($\gamma\approx142$) and $\tau_a=10$ ps
($\tau_a\approx T_\text{rot}/4$).} \label{Pulse}
\end{figure}
\end{center}

Practical generation of such pulses could be done directly by
applying a short pulse right afterward a truncated adiabatic
pulse. Such truncated adiabatic pulses, as produced in
\cite{Stolow2003}, remain however of quite low intensity with the
current technology. An alternative implementation that can be
achieved with state-of-the-art technology consists in spectrally
shaping the phase and the amplitude of a femtosecond laser pulse.
Optimization based on spectral phase shaping has been already
shown to enhance the degree of alignment of the O$_2$ and N$_2$
molecules \cite{HRGLF,RHLF}. The main feature resulting from this
optimization procedure is roughly a sigmoidal phase shape that
corresponds approximately in the time domain to a ramp followed by
a kick. Making the ramp adiabatic or partially adiabatic, as
prescribed above, for a wide range of molecules will require in
general to shape both spectral phase and amplitude of an intense
femtosecond laser pulse. We emphasize that a partially adiabatic
ramp, of duration shorter than for the adiabatic ramp, is of
practical interest in such a shaping technique, since it will be
easier to produce from a femtosecond laser. Figure 4 describes
such a shaping to implement the partially adiabatic ramp followed
by the kick in the conditions of the point marked by ``x'' in Fig.
3 for the CO$_2$ molecule. We have chosen here a Gaussian ramp (of
full width at half maximum $\tau_a=10$ ps and peak intensity
$I_a=2.5$ TW/cm$^2$) and a short pulse (of full width at half
maximum $\tau_k=200$ fs and peak intensity $I_k=25$ TW/cm$^2$)
arriving at the maximum of the Gaussian ramp.
Figure 4 shows the corresponding spectral amplitude and phase,
respectively denoted $A(\omega)$ and $\phi(\omega)$, and defined
through the complex field as ${\cal
E}\sqrt{\Lambda(t)}e^{i\varphi(t)}=\frac{1}{2\pi}\int_{-\infty}^{+\infty}
d\omega A(\omega) e^{i(\omega t+\phi(\omega))}$. Here $\varphi(t)$
is the instantaneous phase, which can be averaged since the
corresponding instantaneous frequency $d\varphi /dt$ is high with
respect to the rotational frequency. These spectral parameters can
be generated for instance from a Gaussian Fourier transform
limited pulse (i.e. corresponding to $\phi(\omega)=0$), centered
at 800 nm with a full width at half maximum of 200 fs, shaped in
phase and amplitude using a spatial light modulators with 640
pixels \cite{Weiner,Chatel}. The field generated taking into
account the pixelization of the device is displayed in the lower
frame of Fig. 4. We estimate that the initial energy required is
about 5 mJ and that the pulse should be focalized on 20 $\mu$m.
With this field, we obtain the expected value
$\max_t\langle\cos^2\theta\rangle\approx 0.992$ in the CO$_2$
molecule at $T=0$ with the complete model (i.e. without
considering the impulsive approximation and taking into account
the centrifugal distortion). A rough estimation of the ionization
probability, using the techniques described in Ref. \cite{Hertz},
gives in this case $P_\text{ion}\lesssim 10^{-4}$. We obtain with
the complete model $\max_t\langle\cos^2\theta\rangle\approx0.915$
for the CO$_2$ molecule at $T=30$ K using $I_a=10$ TW/cm$^2$ and
$I_k=50$ TW/cm$^2$ corresponding to $P_\text{ion}\lesssim 5\times
10^{-3}$.


In conclusion, we have presented a strategy combining an adiabatic
ramp with an impulsive laser field yielding unprecedent efficiency
for the degree of alignment using non-destructive fields. This
strategy can be extended to orient polar molecules. We can use an
adiabatic ramp of an electric field followed by a Half cycle pulse
(HCP). The ramp can be in practice generated for instance by using
the tail of the subsequent HCP itself, or better, by stretching
another HCP. We obtain $\max_t\langle\cos\theta\rangle\approx
0.95$ at $T=0$ with a HCP of peak amplitude 140 kV/cm and of full
width at half maximum $\tau_k=1$ ps, and a ramp of amplitude 2
kV/cm and duration $\tau_a=18$ ps for the KCl molecule
($T_\text{rot}=128$ ps). The implementation of this strategy with
hybrid pulses (i.e. combining a laser and an electric field
\cite{hybrid}) would allow the enhancement of the orientation.
Such a strategy can also be applied to enhance 1-dimensional and
3-dimensional alignment (using shaped elliptical pulses for
instance) of asymmetric tops.

\vfill

\begin{acknowledgments}
We acknowledge support from the Agence Nationale de la Recherche
(ANR CoMoC) and the Conseil R\'{e}gional de Bourgogne.
\end{acknowledgments}


\begin{references}

\bibitem{Seideman_review} H. Stapelfeldt and T. Seideman,
Rev. Mod. Phys. {\bf 75}, 543 (2003).

\bibitem{Stap}
M.D. Poulsen, T. Ejdrup, H. Stapelfeldt, E. Hamilton, and T.
Seideman, Phys. Rev. A \textbf{73}, 033405 (2006).

\bibitem{Vrakking}
C. Siedschlag, O.M. Shir, Th. B\"ack, M.J.J. Vrakking, Opt.
Commun. \textbf{264}, 511 (2006).

\bibitem{DeNalda}
C. Horn, M. Wollenhaupt, M. Krug, T. Baumert, R. de Nalda, and L.
Ba$\tilde{\text{n}}$ares, Phys. Rev. A \textbf{73}, 031401(R)
(2006).

\bibitem{HRGLF} E. Hertz, A. Rouz\'ee, S. Gu\'erin, B. Lavorel, and O.
Faucher, Phys. Rev. A {\bf 75}, 031403(R) (2007).

\bibitem{Rabitz}
M. Leibscher, I.S. Averbukh, and H. Rabitz, Phys. Rev. Lett.
\textbf{90}, 213001 (2003); M. Leibscher, I.S. Averbukh, and H.
Rabitz, Phys. Rev. A \textbf{69}, 013402 (2004).

\bibitem{reaching} D. Sugny, A. Keller, O. Atabek, D. Daems, C.M. Dion,
S. Gu\'erin and H.R. Jauslin, Phys. Rev. A {\bf 69}, 033402
(2004).

\bibitem {Dion}C.M. Dion, A. Keller, and O. Atabek,
Phys. Rev. A, \textbf{72}, 023402 (2005).

\bibitem{Seideman} Z.-C. Yan and T. Seideman,
J. Chem. Phys. {\bf 111}, 4113 (1999); T. Seideman, \textit{ibid.}
{\bf 115}, 5965 (2001).

\bibitem{Stolow2003} J.G. Underwood, M. Spanner, M.Yu. Ivanov,
J. Mottershead, B.J. Sussman, and A. Stolow, Phys. Rev. Lett. {\bf
90}, 223001 (2003).

\bibitem{Fried95} B. Friedrich and D. Herschbach, Phys. Rev. Lett. {\bf %
74}, 4623 (1995).

\bibitem{RHLF} A. Rouz\'ee, E. Hertz, B. Lavorel, and O.
Faucher, J. Phys. B (to be published).

\bibitem{Weiner} A. M. Weiner, Rev. Sci. Inst. {\bf 71}, 1929, (2000).

\bibitem{Chatel} A. Monmayrant and B. Chatel, Rev. Sci. Inst. {\bf 75},
2668, (2004).

\bibitem{Hertz} V. Loriot ,  E. Hertz, A. Rouz\'ee, B. Sinardet, B. Lavorel
and O. Faucher, Opt. Lett. \textbf{31}, 2897 (2006).

\bibitem {hybrid}D. Daems, S. Gu\'erin, D. Sugny and H.R. Jauslin,
Phys. Rev. Lett. \textbf{94}, 153003 (2005).

\end{references}
\end{document}